\documentstyle[sprocl]{article}

\input{psfig}
\def\be{\begin{equation}}
\def\ee{\end{equation}}
\def\bea{\begin{eqnarray}}
\def\eea{\end{eqnarray}}

\begin{document}

\title{Galaxy tracers in cosmological N-body simulations}

\author{ S. GOTTL\"OBER\footnote{
  {\em e-mail:} {\tt sgottloeber@aip.de}. This work was done in 
  collaboration with Anatoly Klypin (Las Cruces), Victor Turchaninov
  (Moscow), and Andrey Kravtsov (Las Cruces) and was supported by the Deutsche
Akademie der Naturforscher Leopoldina with means of the
Bundesministerium f\"ur Bildung und Forschung grant LPD 1996. }
}

\address{Astrophysical Institute Potsdam,
An der Sternwarte 16, D-14482 Potsdam, Germany}

\maketitle\abstracts{Three halo finder algorithms - hierarchical
friends-of-friends, bound density maxima, and 6D minimum spanning tree
- are discussed.  }
  
\section{Introduction}

Galaxies, and the large-scale structure that they trace, grew via
gravitational instability from small amplitude Gaussian density
fluctuations. This nonlinear process can be studied by means of
$N$-body simulations. The main target of this approach is to evaluate
various cosmological theories of structure formation.  Unfortunately,
the dynamical range of current cosmological simulations is inadequate
to follow the details of such a complex process as galaxy formation in
a cosmological environment.  One hopes to find the likely sites where
galaxies form and the most important properties of these
galaxies. However, due to ''overmerging'' $N$-body simulations have
been consistently failing to produce galaxy-size dark matter halos in
dense environments typical for galaxy groups and clusters. The force
and mass resolution required for a simulated halo to survive in galaxy
groups and clusters is extremely high and was not yet reached in these
simulations.  Using the high-resolution Adaptive Refinement Tree (ART)
$N$-body code (Kravtsov et. al. 1997) we followed the evolution of
$\approx 2\times 10^6$ dark matter particles with dynamic range in
spatial resolution of $32,000$.  In these simulations the halos do
survive in regions that would appear overmerged with lower force
resolution. The halo identification in very dense cluster
environments still remains a challenge even with the dynamic range this
high.

\section{Finding galaxies in DM-simulations}

The most widely used halo-finding algorithms are the
friends-of-friends and the spherical overdensity algorithm, however
both are not acceptable (Gelb \& Bertschinger 1994, Summers
et. al. 1995). The friends-of-friends algorithm tends to merge
apparently distinct halos if the linking length is large. Obviously,
this is the case inside of large halos corresponding to groups or
clusters of galaxies.  We developed a ``hierarchical
friends-of-friends''algorithm, which uses a set of different linking
lengths in order to find both isolated halos and halos which appear as
substructures of a larger halo. Applying a bound density algorithm on
the identified halos one can assign galaxies with certain properties
to these halos.

\subsection{Hierarchical friends-of-friends algorithm}

We use a friends-of-friends algorithm with increasing linking length
starting from a small value $l_{vir}/8$ and proceeding by doubling the
linking length: $l =l_{vir}/4$, $l =l_{vir}/2$, and $l=l_{vir}$, where
$l_{vir}$ is the linking radius corresponding to the ``virial
overdensity''.  To avoid huge computational time we derive at first
the minimum spanning tree for the given particles set, from which
clusters at any linking length can be easily obtained. At each level
of the hierarchy every identified cluster of particles (halo
candidate) is marked if none of its particles belongs to a marked
higher-level cluster.  We find a lot of subclumps on all levels in the
large halos, defined on the lowest level of the hierarchy $l_{vir}$,
for example, about 70 (most of them nearby the center and therefore
invisible in projection) in the halo shown in Fig. 1, left.

However, some of these halos are not stable. To identify the real ones,
we run the friends-of-friends algorithm with the same set of linking
lengths using positions of particles at $z=1$.  We identify ``old
halos'' which have at least one progenitor at $z=1$ with galaxies
whereas the others are removed as statistical fluctuations (for
detailed description of the algorithm, see Klypin et al. 1997). Note,
that this approach of excluding halos is important in dense
environments; as a rule all single isolated halos at $z=0$ existed
already at $z=1$.

\subsection{Bound density maxima algorithm}

The bound density maxima algorithm stems from the DENMAX algorithm
(Bertschinger \& Gelb 1991, Gelb \& Bertschinger 1994).  It can work
by itself or in conjunction with the hierarchical
friends-of-friends. In the latter case it starts from the positions
found by the hierarchical friends-of-friends algorithm and removes
unbound particles inside the halo radius, i.e. those particles the
velocity of which is larger than the escape velocity at the position
of the particle. The removal of unbound particles and gentle handling
of halos is important in the case when a halo with a small internal
velocity dispersion moves inside a large group (for detailed
description of the algorithm, see Klypin et al. 1997).

\subsection{6D minimum spanning tree}

Motivated by the observation that a friends-of-friends algorithm often
links particles to a halo which in fact belong to the background and
move in completely other direction,  we have tried to include 
velocity informations into our analysis from the very beginning. 
Thus, we have extended the friends-of-friends algorithm to the
six-dimensional space of particle coordinates and velocities. The main
problem is, how to relate distances in velocity space to coordinate
space. We have defined the distance as
\be ds^2 = dx^2 + dy^2 + dz^2 + \frac{\sigma_x^2}{\sigma_v^2} (dv_x^2
+ dv_y^2 + dv_z^2), 
\ee 
where $\sigma_x$ and $\sigma_v$ are the coordinate and velocity
dispersions of the particles in the object under consideration (for
example a halo). As in the three-dimensional case, at first the
minimum spanning tree has been constructed. In Fig. 2 we show those
particles of the halo presented in Fig. 1 which are connected by the
10,000 shortest links (left for the 3D minimum spanning tree, right
for the 6D minimum spanning tree). The 6D formalism seems to pick up
substructures easier than the 3D formalism. Many of the substructures
found by the hierarchical friends-of-friends algorithm can be seen by
naked eye in Fig. 2, right.

\section{Conclusion}

The halo finding algorithm is a very important tool to study galaxy
formation in high resolution numerical simulations. The
hierarchical friends-of-friends algorithm and the bound density maxima
algorithm have been already successfully used to find stable
halos in $N$-body ART simulations and to determine the properties of
the associated galaxies within CHDM and $\Lambda$CDM cosmological
models (Klypin et al. 1997). The 6D tree algorithm
led to first promising results.

\vskip 2cm

\centerline {\bf Figures}

\bigskip

Fig. 1: $\Lambda$CDM model, 15 $h^{-1}$ Mpc, $128^3$ particles,
resolution $\approx$ 1 kpc, a halo with 154723 particles (7.4\% of 
the total mass in the box) on virial overdensity at $z = 0$ (left), the
same particles at $z=1$ (right). 

\bigskip

Fig.2: 10,000 particles of the halo shown in Fig. 1 ($z=0$) with 
largest density in coordinate space (left) and six-dimensional space of 
coordinates and velocities (right).

\bigskip

{\it Figures are submitted separately as Fig*.gif.}

\end{document}